\documentclass{ap-jnmp}

\copyrightauthor{}

\title{Exact solutions of a nonlinear diffusion equation with absorption and production}

\author{Robert Conte}

\address{Universit\'e Paris-Saclay, ENS Paris-Saclay, CNRS, Centre Borelli, 
\\ F-94235, Cachan, France.\\
Department of mathematics,
The University of Hong Kong, Pokfulam road, Hong Kong.
\\ ORCID https://orcid.org/0000-0002-1840-5095
\\ \today
\\ \email{Robert.Conte@cea.fr}}

\def \ccomma{\raise 2pt\hbox{,}} 

\def \tanh   {\mathop{\rm tanh}\nolimits}

\def \D {\hbox{d}}

\def \kfone {\omega_1}
\def \kkone {\Omega_1}
\def \ka    {\omega_4}
\def \kt    {\Omega}
\def \ex {e_0}

\def\today{29~June~2020} 
\def\today{25~August~2020} 

\begin{document}

\maketitle

\thispagestyle{empty}

\vphantom{\vbox{%
\begin{history}
\received{(30 June 2020)}
\revised{(Day Month Year)}
\accepted{(19 August 2020)}
\end{history}
}}

\begin{abstract} 
We provide closed form solutions for an equation which describes
the transport of turbulent kinetic energy
in the framework of a turbulence model with a single equation.
\end{abstract}


\ccode{2020 Mathematics Subject Classification: 
%
 34A05, 
 35C05, 
 76F60  
}



\section{Introduction}

The partial differential equation (PDE)
\cite{PHM2001}
\begin{eqnarray}
& & {\hskip -18.0 truemm}
\begin{array}{ll}
\displaystyle{
x \hbox{ real},\ 
t>0:\
v_t + \left(v^m\right)_{xx} + c v^p - d v^q=0,\ m>1,\ p>0,\ c>0,\ q>0,\ d>0,
}\\ \displaystyle{
v(x,0)=v^0(x) \ge 0 \hbox{ with compact support},
}
\end{array}
\label{EDPv}
\end{eqnarray}
governs the transport of turbulent kinetic energy $k$
in the framework of a turbulence model with a single equation.
This minimal model retains the essential physical ingredients:
time evolution ($\partial_t$),
space diffusion ($\partial_x^2$),
one space dimension only,
nonlinearity ($m\not=1$),
dissipation ($c v^p$) 
and production ($d v^q$).
Our goal in the present paper is to find closed form solutions
to serve as a validation test for the numerical schemes.
In order to achieve that,
we will not design new methods
but apply existing methods to generate several new solutions
of physical interest.

The so-called ``model equation'' \cite{PHM2001}
corresponds to
$m = p = 3/2 , c = d = 1, q= 1/2$,
and the very special case $c=d=0$  (of no interest to us)
is known as the porous media equation \cite{Aronson,Zeldovich_Raizer}.
\medskip

In order to search for closed form solutions,
it is better to first convert (\ref{EDPv}) to an algebraic equation,
\begin{eqnarray}
& & {\hskip -18.0 truemm}
v^{m-1}=w,\
w_t - m w w_{xx} -\frac{m}{m-1} {w_{x}}^2 
+(m-1) c w^{\displaystyle\frac{p+m-2}{m-1}} 
-(m-1) d w^{\displaystyle\frac{q+m-2}{m-1}} 
=0,
\end{eqnarray}
with the choice $p=m$, $q=2-m$, $m$ arbitrary
(since it contains the case of physical interest $p=3/2$, $q=1/2$ when $m=3/2$).

Therefore the final algebraic equation which we will study is (after rescaling $t$)
\cite[Eq.~(3.24)]{PHM2001}
\begin{eqnarray}
& & {\hskip -18.0 truemm}
\lambda v^{m-1}=w:\ E(w) \equiv 
 w_t - b w w_{xx} - {w_{x}}^2 + c w^2 - d=0,\ b>0,\ c>0,\ d>0.
\label{EDPw}
\end{eqnarray}
Since there could exist physical systems not requiring the positivity of $b,c,d$,
we will also mention a few solutions with $b,c,d$ not all positive.

\bigskip
The paper is organized as follows.
In section \ref{sectionGP} we first recall 
the ingenious, however not generalizable, method 
which has allowed Maire to obtain a solution matching all the physical constraints
in (\ref{EDPw}).

Section \ref{sectionSymetries} is devoted,
using the method of infinitesimal Lie point symmetries,
to the construction of 
all the reductions of the PDE (\ref{EDPw})
to
an ordinary differential equation (ODE),
then to their integration.

In section \ref{sectionSingularites},
we study the \textit{local} behaviour of the field $w(x,t)$
near its movable singularities,
a prerequisite to the search for closed form solutions.

The next section \ref{section-One-family-truncation}
is devoted to a search for new exact solutions
based on the singularity structure.

\section{Method of Galaktionov and Posashkov}
\label{sectionGP}

The PDE (\ref{EDPw}) belongs to the class
\begin{eqnarray}
& & w_t - P(\partial_x,w)=0,\ P \hbox{ differential polynomial},
\label{eqPDE-GP}
\end{eqnarray}
with the additional property of existence of a finite number
of functions $f_k(x)$ whose linear
combinations are stable under $P(\partial_x,w)$,
\begin{eqnarray}
& & {\hskip -8.0 truemm}
\forall (\alpha_1, \dots, \alpha_K)\
 \exists (\beta_1, \dots, \beta_K):\
P(\partial_x, \sum_{k=1}^K \alpha_k f_k(x))=\sum_{k=1}^K \beta_k f_k(x).
\label{eqPDE-GP-stable}
\end{eqnarray}
\vspace{0.2cm}
Then Galaktionov and Posashkov \cite{GP} observed that
the (kind of ``adiabatic'') assumption
\begin{eqnarray}
& & 
w(x,t)=\sum_{k=1}^K \gamma_k(t) f_k(x),
\end{eqnarray}
amounts to solving a nonlinear system of ODEs (no more PDEs) 
for the functions $\gamma_k(t)$.
In the present case (\ref{EDPw}),
\begin{eqnarray}
& & w(x,t)=\gamma_1(t) + \gamma_2(t) \cosh(k x),\ 
\label{eq-GP-class}
\nonumber
\end{eqnarray}
one thus obtains four solutions leaving the physical parameters $b,c,d$ unconstrained
and positive.

The most physically relevant solution
\cite[Eq.~(3.37)]{PHM2001},
\begin{eqnarray}
& &
\left\lbrace 
\begin{array}{ll}
\displaystyle{
\lambda v^{m-1}=w=A \frac{\cosh(\omega t) -\cosh(k x)}{\sinh(\omega t)},\
}\\ \displaystyle{
\omega=A c \frac{2+b}{1+b},\ k^2=\frac{c}{1+b},\ A^2=\frac{d}{c},\
}
\end{array}
\right.
\label{eq-w-sol-PHM}
\end{eqnarray}
does not depend on any movable
(i.e.~function of the initial conditions) constant,
excluding of course the arbitrary origins $x_0$ and $t_0$,
and it provides a good description \cite{PHM2001}
of the transport of the turbulent kinetic energy.

The second solution is stationary,
\begin{eqnarray}
& & 
w=A \cosh (k x),\  k^2=\frac{c}{1+b},\ A^2=\frac{(1+b) d}{c},\
\label{eq-w-sol-coshx}
\end{eqnarray}
and the third one homogeneous in space
\begin{eqnarray}
& & {\hskip -14.0 truemm}
w=\frac{\omega_2}{2 c} \tanh \frac{\omega_2}{2} t,\ \omega_2^2=4 c d.
\label{eqsoltrunc1wa}
\label{eq-w-sol-tanht}
\end{eqnarray} 

Finally, the fourth solution is characterized by a second order nonlinear ODE,
\begin{eqnarray}
& &
\left\lbrace 
\begin{array}{ll}
\displaystyle{
w=g_0(t)+g_1(t) \cosh(k x),\ k^2=\frac{c}{1+b},\
g_1^2=g_0^2-\frac{d}{c}\not=0,\
}\\ \displaystyle{
-(b+1) g_0'' - 2 (2 b+3) c g_0 g_0' -2(b+2) c g_0 (c g_0^2-d)=0,
}
\end{array}
\right.
\label{eq-w-sol-odeg0}
\end{eqnarray}
and, if one excludes the particular values $g_0$ listed in the three previous solutions,
its only physical ($b$ positive) solutions are multivalued \cite[Eq.~(9.2)]{BureauMI},
but nevertheless expressible \textit{via} quadratures \cite{LemmerLeach}.
The same kind of ODE will again be encountered hereafter, see Eq.~(\ref{eqTrunc1system}) for $F_1$.
\medskip

If one relaxes the constraint $b>0$,
there also exists a fifth solution,
\begin{eqnarray}
 & & {\hskip -14.0 truemm}
w=\frac{\omega_2}{2 c} \tanh \frac{\omega_2}{2} t + A \cos (k x),\ 
\omega_2^2=4 c d + 4 c^2 A^2,\ 
k^2=c,\ 
b=-2,\ 
A=\hbox{arbitrary}.
\label{eqsoltrunc1wb}
\label{eq-w-sol-tanht-cosx}
\end{eqnarray}

Since the above ingenious method only applies to the restricted class 
(\ref{eqPDE-GP})--(\ref{eqPDE-GP-stable}),
let us in addition 
apply the two main classes of methods able to find explicit solutions of algebraic PDEs~:
\begin{itemize}

\item 
Those based on the symmetries of the PDE,
which generate reductions to ODEs;

\item 
Those based on the movable singularities of the PDE,
which (after a double study, local then global)
generate closed form solutions $w(x,t)$.

\end{itemize}

\section{Solutions obtained by symmetries}
\label{sectionSymetries}

For generic values of $b,c,d$, the only Lie point symmetries of the PDE (\ref{EDPw}) 
are arbitrary translations of both $x$ and $t$.
The resulting characteristic system,
\begin{eqnarray}
& & 
\frac{\D x }{\alpha}=\frac{\D t}{\beta}=\frac{\D w}{0},\
\alpha,\beta \hbox{ arbitrary constants},
\end{eqnarray}
admits the two invariants $w$ and $\beta x - \alpha t$,
with $\alpha,\beta$ not both zero,
thus defining the unique reduction to an ODE,
\begin{eqnarray}
& & {\hskip -18.0truemm}
w(x,t)=W(\xi),\ \xi=\beta x - \alpha t,\
-\alpha W' - \beta^2 (b W W'' +{W'}^2) + c W^2 -d=0.
\label{EDOW}
\end{eqnarray}
One must distinguish $\beta\not=0$
(the reduction preserves the differential order two)
and $\beta=0$ (the reduction lowers it to one).

For $\beta\not=0$,
since the positive parameter $b$ is never equal to $-(1-1/n)$,
with $n$ some signed integer,
the general solution $W(\xi)$ is multivalued \cite{PaiBSMF}
and generically cannot be obtained in closed form.
A notable exception is $\beta\not=0$ and $\alpha=0$
(stationary wave),
\begin{eqnarray}
& & 
\xi=x,\ w(x,t)=W(x),\ - b W W'' -{W'}^2 + c W^2 -d=0,
\label{EDOx}
\end{eqnarray}
which admits the first integral \cite{PHM2001}
\begin{eqnarray}
& & 
K=\left[(1+b)({W'}^2-d)+c W^2 \right] W^{2/b}.
\label{EDOZ}
\end{eqnarray}
For $K=0$, the general solution is physically acceptable
and has already been obtained, see Eq.~(\ref{eq-w-sol-coshx}), 
and for $K\not=0$ the solution is given implicitly by the quadrature
\begin{eqnarray}
& & 
w(x,t)=W(x),\
x=x_0 + \int 
\left(\frac{1+b}{(1+b) d - c W^2 - K W^{-2/b}} \right)^{1/2} \D W.
\label{eq-w-sol-quadrature}
\end{eqnarray}
The only cases of invertibility of this quadrature are
$-2/b=1,2,3,4$,
yielding expressions $W(x)$ trigonometric $(b=-2,-1)$
or elliptic ($b=-2/3,-1/2$),
which however all violate the physical requirement $b>0$.
\medskip

For $\beta=0$ one obtains the front independent of $x$, Eq.~(\ref{eq-w-sol-tanht}).
				
\section{Structure of singularities}
\label{sectionSingularites}

Any closed form solution depends on arbitrary functions or constants,
such as $x_0,t_0$ in (\ref{eq-w-sol-PHM}),
which may define movable singularities.
For instance, the solution (\ref{eq-w-sol-PHM}) definies
two families of movable singularities~:
on one side the movable poles of $w$
located at the points $t=t_0+ n i \pi/\omega, n \in \mathbb{Z}$,
on the other side movable poles of $1/w$ (movable zeroes of $w$)
located on the \textit{singular manifold} $\varphi(x,t)=0$ defined by
\begin{eqnarray}
& & 
\varphi(x,t) \equiv \cosh\omega (t-t_0) -\cosh k (x-x_0)=0.
\end{eqnarray}
\smallskip

A prerequisite to the systematic search for solutions
is therefore the determination of the structure of 
the movable singularities of (\ref{EDPw}),
\begin{eqnarray}
& & 
w \sim_{\varphi \to 0} w_0 \varphi^p,\ p \notin \mathcal{N}.
\end{eqnarray}
Since the highest derivative term $w w_{xx}$ displays
the singularity $w=0$,
one must also study the movable zeroes of $w$ (movable poles of $w^{-1}=f$),
\begin{eqnarray}
& & {\hskip -8.0truemm}
\lambda v^{m-1}=w=f^{-1}:\ E(f)\equiv -f^2 f_t + b f f_{xx}-(2 b +1)f_x^2 - d f^4 +c f^2=0.
\label{EDPf}
\end{eqnarray} 
This is a classical computation \cite[\S 4.4.1]{CMBook},
whose results are the following.

The PDE (\ref{EDPw}) admits two types of movable singularities.
\begin{enumerate}
\item
If $\varphi_x=0$ (the singlar manifold is then said characteristic),
the highest derivative term $w w_{xx}$ does not contribute
to the leading order,
and $w$ (as well as $f$) presents one family of movable simple poles, 
\begin{eqnarray}
& & 
\varphi_x=0:\ w \sim_{\varphi \to 0} c^{-1} \varphi_t \varphi^{-1},\
\label{eq-family-char-w}
\\ & &
\varphi_x=0:\ w^{-1} \sim_{\varphi \to 0} d^{-1} \varphi_t \varphi^{-1}.
\label{eq-family-char-f}
\end{eqnarray}

\item
If $\varphi_x\not=0$ (noncharacteristic singlar manifold),
$w$ has no movable poles
and $w^{-1}=f$ presents two families of movable simple poles, 
\begin{eqnarray}
& & {\hskip -8.0truemm}
f \sim_{\varphi \to 0} f_0 \varphi_x \varphi^{-1},\ 
d f_0^2 - \frac{\varphi_t}{\varphi_x} f_0 +1=0,\ f_0 \not=0.
\label{eq-family-nonchar-f}
\end{eqnarray}

\end{enumerate}

To finish this local analysis,
one must then compute the Fuchs indices of
the linearized equation of (\ref{EDPw})
in the neighborhood of $\varphi=0$.
Indeed, a necessary condition of singlevaluedness is that all Fuchs indices be integers
of any sign.
For the families (\ref{eq-family-char-w}) or (\ref{eq-family-char-f}),
the unique Fuchs index is $i=-1$.
For each of the two families (\ref{eq-family-nonchar-f}),
one finds
\begin{eqnarray}
& & {\hskip -8.0truemm}
i=-1,\ \frac{d f_0^2 -1}{b},
\label{eqFuchsindex}
\end{eqnarray}
and the noninteger value (even nonrational) of the second index
reflects the high level of nonintegrability of the initial PDE.
\medskip

In order to build solutions of this kind of nonintegrable PDE,
the various methods based on the singularity structure
are reviewed in summer school lecture notes \cite{CetraroConte},
where the proper original references can be found.
Let us now investigate a few of them.
		
\section{Truncations}
\label{section-One-family-truncation}

They consist in requiring the Laurent series 
of a single family to terminate.
For the respective local behaviours
(\ref{eq-family-char-w}), (\ref{eq-family-char-f}) and (\ref{eq-family-nonchar-f}),
the corresponding possible solutions are defined by
\begin{eqnarray}
& & {\hskip -8.0truemm}
(\varphi_x=0):\ w = c^{-1} \chi^{-1} + w_1,
\label{eq-one-family-truncation-w}
\\
& & {\hskip -8.0truemm}
(\varphi_x=0):\ f = d^{-1} \chi^{-1} + f_1,
\label{eq-one-family-truncation-f}
\\
& & {\hskip -8.0truemm}
(\varphi_x\not=0):\ f=f_0 \chi^{-1} + f_1,\ f_0 \not=0,
\label{eq-one-family-nonchar-truncation-f}
\end{eqnarray}
in which the expansion variable $\chi(x,t)$ 
is any homographic transform of $\varphi(x,t)$
vanishing with $\varphi$.
There exists an optimal choice of $\chi$ \cite{Conte1989},
characterized by its gradient
($(\xi,\eta)$ represents either $(x,t)$ or $(t,x)$)
\begin{eqnarray}
& &
\chi_\xi= 1 + \frac{S}{2} \chi^2,\
\chi_\eta=- C + C_\xi \chi  - \frac{1}{2} (C S + C_{\xi\xi}) \chi^2,
\label{eqChix}
\label{eqChit}
\end{eqnarray}
and the constraint
\begin{eqnarray}
& &
S_\eta + C_{\xi\xi\xi} + 2 C_\xi S + C S_\xi = 0.
\label{eqCrossXT}
\end{eqnarray}

After substitution in equations (\ref{EDPw}) and (\ref{EDPf}),
the LHS $E$ of these equations become Laurent series in $\chi$ which also terminate,
\begin{eqnarray}
& & {\hskip -8.0truemm}
E \equiv \sum_{j=0}^{-q} E_j \chi^{j+q},\ -q \in \mathbb{N},
\end{eqnarray}
and the coefficients $E_j$ only depend on $w_0,w_1$ (or $f_0,f_1$) and $(S,C)$.
The solutions are then provided by solving the determining equations
\begin{eqnarray}
& & {\hskip -8.0truemm}
\forall j=0,...,-q:\ E_j(w_0,w_1,S,C)=0.
\end{eqnarray}

\subsection{Characteristic one-family truncation of $w$}

The truncation (\ref{eq-one-family-truncation-w}) with $\chi_x=0$
defines the system
\begin{eqnarray}
& &
\left\lbrace 
\begin{array}{ll}
\displaystyle{
E_0 \equiv 0,\ 
}\\ \displaystyle{
E_1 \equiv -b c^{-1} w_{1,xx} + 2 w_1=0,\ 
}\\ \displaystyle{
E_2 \equiv -b w_1 w_{1,xx} - {w_{1,x}}^2 + w_{1,t} + c w_1^2 - \frac{S}{2 c} - d=0,
}
\end{array}
\right.
\end{eqnarray} 
whose general solution is
\begin{eqnarray}
& &
w_1=A \cos (k x),\
S=-2 c d-2 c^2 A^2,\
k^2=c,\ 
(b+2) A=0.
\end{eqnarray} 
The value of $\chi$ results by integrating the Riccati equation (\ref{eqChix}),

\begin{eqnarray}
& &
\chi^{-1}=\frac{\omega_2}{2} \tanh \frac{\omega_2}{2} t,\ \omega_2^2=4 c d + 4 c^2 A^2.
\end{eqnarray}
The bifurcation $(b+2) A=0$ defines two previously found solutions $w$, 
(\ref{eq-w-sol-tanht}) and
(\ref{eq-w-sol-tanht-cosx}).

\subsection{Characteristic one-family truncation of $f$}

The truncation (\ref{eq-one-family-truncation-f}) with $\chi_t=0$
defines the system
\begin{eqnarray}
& &
\left\lbrace 
\begin{array}{ll}
\displaystyle{
E_0 \equiv 0,\ 
}\\ \displaystyle{
E_1 \equiv f_1=0,
}\\ \displaystyle{
E_2 \equiv c d + \frac{S}{2} - 5 d^2 f_1^2 - d f_{1,t}=0,
}\\ \displaystyle{
E_3 \equiv 2 c d f_1 + S f_1 -4 d^2 f_1^3 -2 d f_1 f_{1,t} + b d f_{1,xx}=0,
}\\ \displaystyle{
E_4 \equiv E(f_1) + \frac{S}{2 d} f_1^2=0,
}
\end{array}
\right.
\end{eqnarray} 
whose general solution is
\begin{eqnarray}
& &
f_1=0,\
S=-2 c d.
\end{eqnarray} 
After integration of the Riccati equation (\ref{eqChix}),
\begin{eqnarray}
& &
\chi^{-1}=\frac{\omega_2}{2} \tanh \frac{\omega_2}{2} t,\ \omega_2^2=4 c d,
\end{eqnarray}
one obtains
\begin{eqnarray}
& & 
f= d^{-1} \frac{\omega_2}{2} \tanh \frac{\omega_2}{2} t,\ \omega_2^2=4 c d,
\end{eqnarray}
identical to  (\ref{eq-w-sol-tanht}).

To conclude, these characteristic truncations yield nothing new.

\subsection{Noncharacteristic one-family truncation of $f$}


Let us finally consider one of the two families 
(\ref{eq-family-nonchar-f}),
whose Fuchs indices are 
(\ref{eqFuchsindex}),
and let us assume $b d \not=0$.
The truncation (\ref{eq-one-family-nonchar-truncation-f})
defines the system
\begin{eqnarray}
& &
\left\lbrace 
\begin{array}{ll}
\displaystyle{
E_0 \equiv d f_0^2 + C f_0 +1=0, 
}\\ \displaystyle{
E_1 \equiv 2 (-2 d f_0^2 - C f_0 + b) f_1 
     + f_0^2 C_x - f_0 f_{0,t} + 2 (1+b) f_{0,x}=0,
}\\ \displaystyle{
E_j \equiv E_j(f_0,f_1,S,C)=0,\ j=2,3,
}\\ \displaystyle{
E_4 \equiv b (2 b+1)
\left[f_0^2 S + 2 f_1^2 + 2 f_1 f_{0,x} - 2 f_0 f_{1,x} \right]^2=0,
}\\ \displaystyle{
X \equiv S_t + C_{xxx} + 2 C_x S + C S_x = 0,
}
\end{array}
\right.
\label{eq-one-family-truncation-f-nonchar}
\end{eqnarray} 
and the factorization of $E_4$ makes the resolution easy.
It is even easier after the change of functions
\begin{eqnarray}
& & {\hskip -8.0truemm}
(f_0,f_1) \to (F_0,F_1):\ 
f_0=1/F_0,\
f_1=f_0 F_1 -\frac{1}{2} f_{0,x}.
\label{eqfjtoFj}
\end{eqnarray}

In the case $b\not=-1/2$,
the algebraic (i.e.~nondifferential) elimination 
of $F_{0,t},F_{1,t},S_x$
yields the much simpler equivalent system,
\begin{eqnarray}
& & {\hskip -10.0truemm}
b\not=-\frac{1}{2}:\
\left\lbrace 
\begin{array}{ll}
\displaystyle{
E_1\equiv F_{0,t} -(b+2) F_0 F_{0,x} +2 (b+1) F_0^2 F_1 - 2 d F_1=0,
}\\ \displaystyle{
E_2 \equiv - b F_{0,xx}
 +2 (b+2) F_1 F_{0,x} 
}\\ \displaystyle{\phantom{1234567}
 +2 [(3 b+2) F_{1,x} -(4 (b+1) F_1^2 + c)] F_0
-2 F_{1,t}=0,
}\\ \displaystyle{
X \equiv  b F_{1,xx} -2(3 b+2) F_1 F_{1,x} + 4 (b+1) F_1^3 - c F_1=0,
}\\ \displaystyle{
S=\frac{F_{0,xx}}{F_0}-\frac{3}{2}\frac{F_{0,x}^2}{F_0^2}
-2 F_1^2 +2 F_{1,x} -2 \frac{F_{0,x}}{F_0} F_1,
}\\ \displaystyle{
C=-F_0-d F_0^{-1}.
}
\end{array}
\right.
\label{eqTrunc1system}
\end{eqnarray} 

This system (\ref{eqTrunc1system}) is solved in three steps:
\begin{description}
\item
(i) integration of the ODE in $F_1$ defined by $X=0$,
which introduces at most two arbitrary functions of $t$;
\item
(ii) determination of $F_0$ by solving the overdetermined system
$(E_1=0,E_2=0)$;
\item
(iii) knowing the values of $(S,C)$,
integration of the Riccati system (\ref{eqChix}) for $\chi$.
\end{description}

In the unphysical case $b=-1/2$,
we could not find an equivalent system as simple as 
(\ref{eqTrunc1system}).

Let us now perform the above mentioned three steps.

\subsubsection{Values of $F_1$}
\label{sectionF1}

As already mentioned about the similar ODE (\ref{eq-w-sol-odeg0}),
we will discard the multivalued solutions of the ODE $X=0$ for $F_1(t)$,
refering the interested reader to Ref.~\cite{LemmerLeach}.

All singlevalued solutions of the ODE for $F_1$ 
are obtained by two methods:
for the general solution,
by looking in the classical exhaustive tables 
\cite{PaiBSMF,GambierThese};
for particular solutions,
by looking for Darboux polynomials.
One thus finds exactly seven solutions,
five of them with a negative $b$
(unphysical for the diffusion problem, but possibly admissible for other systems)
and two with an arbitrary value of $b$,
\begin{eqnarray}
b=-\frac{4}{3}&:& F_1=-\partial_x \log\left[\cosh(k x -g(t))-\cosh(K(t))\right],\ k^2=-\frac{3}{4}c,\
\label{eqF1listfirst}
\\
b=-\frac{4}{5}&:& F_1=-\partial_x \log\left[\wp(x-g(t))-\ex \right],\
\ex=-\frac{5 c}{48},\
g_2=12 \ex^2,\ g_3(t),
\label{eqF145}
\\
b=-\frac{1}{2}&:& F_1=\frac{1}{2} \partial_x \log\left[\wp(x-g(t))-\ex \right],\
\ex=\frac{c}{6},\
g_2=12 \ex^2,\ g_3(t),
\\
b=-\frac{2}{3}&:& F_1=\sqrt{\wp(x-g(t))-\ex},\ 
\ex=-\frac{c}{2},\
g_2=12 \ex^2 - 4     K(t),\ g_3=-8 \ex^3 + 4 \ex K(t),
\label{eqF1listlastb}
\\
b+1\not=0&:& F_1=-                 \frac{k}{2} \tanh\frac{k}{2} (x -g(t)),\ k^2=\frac{c}{b+1},
\label{eqF1tanh1}
\\
b+1\not=0&:& F_1=- \frac{b}{2(b+1)}\frac{k}{2} \tanh\frac{k}{2} (x -g(t)),\ k^2=\frac{4(b+1) c}{b^2},
\label{eqF1tanh2}
\\
b=-2     &:& F_1=0,
\label{eqF10}
\end{eqnarray}
in which $\wp(x,g_2,g_3)$ is the elliptic function of Weierstrass,
and $g$, $K$ two arbitrary functions of $t$.
The first four depend on two arbitrary functions of $t$,
the next two on one arbitrary function of $t$.

For $b=-1/2$, the obtained solution is a particular case of the solution,
which we could not obtain,
resulting from the system (\ref{eq-one-family-truncation-f-nonchar}).

For $b=-1$ and $c\not=0$, the ODE for $F_1$ has no singlevalued solution.

\medskip
Let us next determine $F_0$.
By the elimination 
of $F_1$ between $E_1=0$ and $E_2=0$,
the value of $F_0$ is the root of a sixth degree polynomial
whose coefficients are polynomial in $F_1$ and its derivatives.
However, this computation is only tractable for the two $\tanh$ solutions 
(which only depend on $g(t)$), ,
and one finds constant values for $F_0$ and $g'(t)$,
\begin{eqnarray}
b+1\not=0&:& F_0^2=\frac{d}{b+1},\ g'(t)=+(b+2) \frac{k}{2} F_0,\ 
\\
b+1\not=0&:& F_0^2=\frac{d}{b+1},\ g'(t)=-(b+2) \frac{k}{2} F_0.\ 
\end{eqnarray}

In the four other cases $b=-4/3, -4/5,-1/2,-2/3$ of the list (\ref{eqF1listfirst})--(\ref{eqF1listlastb}),
this technical difficulty is overcome 
by integrating the linear inhomogeneous ODE $E_2=0$ for $F_0$ as follows.

One first notices that, in its homogeneous part,
the simple pole of $F_1$ with residue $r=-1$
is a Fuchsian singularity for the EDO in $F_0$,
whose Fuchs indices $i$, the roots of
\begin{eqnarray}
& & b i^2 -(b+4 r +2 b r) i +8 (b+1) r^2 +2 (3 b+2) r=0,
\end{eqnarray}
are irrational for the four values of $b$.
Since $F_0$ is necessarily an algebraic function of $F_1$ and its derivatives,
the only such algebraic solution of the homogeneous part of 
$E_2=0$ is $F_0=0$,
see the example Eq.~(\ref{eq43F0brut}) hereafter.
One then computes $F_0$ as a particular solution of the 
inhomogeneous equation $E_2=0$.
Once this value of $F_0$ obtained,
the nonlinear equation $E_1=0$
generates constraints on $g(t)$ and $K(t)$.
This method is exemplified in section \ref{section43}.

\subsubsection{Solution, case of the first $\tanh$ value of $F_1$}

Given
\begin{eqnarray}
b+1\not=0&:& F_0=\frac{1}{a_0},\ a_0^2=\frac{b+1}{d},\ g'(t)=+(b+2) \frac{k}{2} F_0,
\end{eqnarray}
one finds successively
\begin{eqnarray} & &
S=-\frac{k^2}{2},\
C=-\frac{b+2}{a_0},\
\end{eqnarray}
then, by integration of the Riccati system (\ref{eqChix})--(\ref{eqChit})
\begin{eqnarray} & &
\chi^{-1}= \frac{k}{2} \tanh \frac{k}{2} (x+\frac{b+2}{a_0} t),
\end{eqnarray}
the solution
\begin{eqnarray} & &
w^{-1}=f=a_0 \frac{k}{2} \left[\tanh \frac{k}{2} (x+\frac{b+2}{a_0} t) 
                             - \tanh \frac{k}{2} (x-\frac{b+2}{a_0} t)\right],\
k^2=\frac{c}{b+1}\ccomma
\end{eqnarray}
which is just another representation of the solution (\ref{eq-w-sol-PHM})
already found by the method of Galaktionov and Posashkov.

\subsubsection{Solution, case of the second $\tanh$ value of $F_1$}
\label{section-second-tanh}

Solving $(E_1,E_2)$ for $(F_0(x,t),g(t))$ is again quite easy,
\begin{eqnarray} {\hskip -5.0truemm}
& &
F_0(x,t)=\hbox{const}=\frac{1}{a_0},\
a_0^2=\frac{b+1}{d},\ 
g(t)=-\frac{(b+2) k}{2 a_0} t,\
\nonumber 
\end{eqnarray}
then one obtains
\begin{eqnarray}
& & {\hskip -10.0truemm}
S=-\frac{b}{b+1} \frac{k^2}{4} + \frac{b(b+2)}{2 (b+1)^2} \frac{k^2}{4} \tanh^2 \frac{k}{2}(x +\frac{b+2}{a_0} t),\
C=-\frac{b+2}{a_0}\cdot
\nonumber 
\end{eqnarray}
The ODE for $\chi^{-1}$ is a Lam\'e equation in its Riccati form,
whose solution is
singlevalued for $b=-1+1/n$, $ n \in \mathbb{Z}$,
multivalued otherwise.
For the present diffusion problem,
this new solution leaves $b,c,d$ unconstrained and can indeed be used to test
the validity of numerical schemes.

\subsubsection{Case $b=-2$}

A computation similar to the above one yields
\begin{eqnarray} 
& &
F_1=0,\
F_0=A \cosh k x,\
S=-c \left[1 - \frac{3}{2} \tanh^2 k x \right],\
C= - \frac{A^2+d}{d} \cosh k x,\
k^2=-c,
\end{eqnarray}
then 
\begin{eqnarray} 
& &
\chi^{-1}=\partial_x \log \left\lbrack  
\cosh(k x)^{-1/2} \left( -\frac{\kt}{A k} \cosh(\kt t) + \sinh(\kt t) \cosh(k x)  \right)
\right\rbrack,\
\kt^2=(A^2+d) c,
\end{eqnarray}
i.e.~a solution identical to (\ref{eq-w-sol-tanht-cosx})
already found in section \ref{sectionGP}.

\subsubsection{Solutions, case $b=-4/3$}
\label{section43}

The resolution of the four cases $b=-4/3, -4/5, -1/2, -2/3$ follows the same pattern,
so we only detail it for $b=-4/3$ (Eq.~(\ref{eqF1listfirst})).

For this value $b=-4/3$, the ODE $E_2=0$ possesses the general solution
\begin{eqnarray}
& & {\hskip -16.0truemm}
F_0=
 g_+(t) \frac{[\cosh K(t) \cosh(k x-g(t))+\sinh K(t) \sinh(k x-g(t))-1]^{ \sqrt{2}}}
  {(\cosh(k x-g(t))-\cosh K(t))^{\sqrt{2}-1}}
\nonumber \\ & & {\hskip -16.0truemm} \phantom{1+}
+g_-(t) \frac{[\cosh K(t) \cosh(k x-g(t))+\sinh K(t) \sinh(k x-g(t))-1]^{-\sqrt{2}}}
  {(\cosh(k x-g(t))-\cosh K(t))^{-\sqrt{2}-1}}
\nonumber \\ & & {\hskip -16.0truemm} \phantom{1+}
+3\frac{\left[
K' \sinh K(t) \sinh(k x-g(t)) + g' \cosh K(t) \cosh(k x-g(t)) -g'
\right]}{2 k \sinh^2 K(t)},
\label{eq43F0brut}
\end{eqnarray} 
in which $g_\pm(t)$ are two other arbitrary functions of $t$.
\smallskip

As already argued in section \ref{sectionF1},
since the relation between $F_0$ and $e^{k x-g(t)}$ is necessarily algebraic,
the two functions $g_+$ and $g_-$ must vanish.
Equation $E_1=0$ then yields the necessary and sufficient conditions,
\begin{eqnarray} 
& &
g''=0,\ K''=0,\ g' K'=0,\ {g'}^2+{K'}^2=c d,\ 
\end{eqnarray}
solved as
\begin{eqnarray} 
& &
g(t)=\kfone t ,\ K(t)=\kkone t + k_0,\ \kfone \kkone=0,\ \kfone^2+\kkone^2=c d,
\end{eqnarray}
in which $\kfone,\kkone,k_0$ are constant.

The system (\ref{eqTrunc1system}) therefore has for solution,
in the first case $\kfone \not=0,\kkone=0$, 
\begin{eqnarray}
& & {\hskip -10.0truemm}
\left\lbrace 
\begin{array}{ll}
\displaystyle{
F_1=-\partial_x \log\left[\cosh(k x - \kfone t)-\cosh k_0\right],\ k^2=-\frac{3}{4}c,\ \kfone^2=c d,
}\\ \displaystyle{
F_0=\frac{3 \kfone }{2 k \sinh^2 k_0} \left[\cosh k_0 \cosh(k x-\kfone t) -1\right],
}\\ \displaystyle{
}\\ \displaystyle{
S=-\frac{k^2}{2}
  \left[1-\frac{3 \sinh^2 k_0}{(\cosh k_0 \cosh(k x-\kfone t) -1)^2}\right],
}\\ \displaystyle{
C=-\frac{\kfone}{2 k}
\frac{3(\cosh k_0 \cosh(k x-\kfone t) -1)^2-\sinh^4 k_0}
     {\sinh^2 k_0 (\cosh k_0 \cosh(k x-\kfone t) -1)},
}
\end{array}
\right.
\end{eqnarray} 
and in the second case $\kfone=0,\kkone \not=0$, 
\begin{eqnarray}
& & {\hskip -10.0truemm}
\left\lbrace 
\begin{array}{ll}
\displaystyle{
F_1=-\partial_x \log\left[\cosh k x-\cosh \kkone t\right],\ 
F_0=\frac{3 \kkone \sinh k x}{2 k \sinh \kkone t},\
k^2=-\frac{3}{4}c,\ \kkone^2=c d,\
}\\ \displaystyle{
S=-\frac{k^2}{2}\left(1+\frac{3}{\sinh^2 k x}\right),\
C=\frac{\kkone}{2 k} 
 \left[\frac{\sinh \kkone t}{\sinh k x}-3 \frac{\sinh k x}{\sinh \kkone t}\right].
}
\end{array}
\right.
\end{eqnarray} 

The integration of the Riccati system (\ref{eqChix})--(\ref{eqChit})
introduces another arbitrary constant $t_0$,
then the solutions $w$ are defined by
(\ref{eq-one-family-nonchar-truncation-f})
and (\ref{eqfjtoFj}).
One thus obtains two solutions $w$,
respectively
\begin{eqnarray}
& & {\hskip -15.0truemm}
\left\lbrace 
\begin{array}{ll}
\displaystyle{
\chi^{-1}=\frac{k}{2}
\frac{a_1  \sinh\xi (\cosh k_0 \sinh \xi + \cosh^2 k_0-2)
     +a_2 (\cosh k_0 \sinh^2 \xi -2 \sinh \xi + \cosh k_0)}
     {[a_1(\sinh \xi-\cosh k_0)+a_2 \sinh\xi]
      [\cosh k_0 \sinh \xi-1]},\
}\\ \displaystyle{
\frac{a_1}{a_2}=\frac{\ka \tanh(\ka (t-t_0))}{2 \kfone},\
\xi=k x -\kfone t,\
k^2=-\frac{3}{4}c,\ 
\kfone^2=c d,\
\ka^2=\left(1-\frac{3}{\sinh^2 k_0}\right) c d,\ 
}\\ \displaystyle{
w=\frac{3 \kfone (\cosh k_0 - \cosh \xi)}{2 k^2 \sinh^2 k_0}
\left[ 
\kfone - \frac{\ka}{2} \tanh(\ka (t-t_0))(\cosh k_0 - \cosh \xi)
\right],\
}
\end{array}
\right.
\label{eqsoltruncwnc43a}
\end{eqnarray} 
and
\begin{eqnarray}
& & {\hskip -10.0truemm}
\left\lbrace 
\begin{array}{ll}
\displaystyle{
\chi^{-1}=\frac{k}{2}\frac{a_1 (\cosh^2 k x -2) - a_2 \cosh k x}
                         {[a_1  \cosh k x + a_2] \sinh k x},\ 
k^2=-\frac{3}{4}c,\ 
}\\ \displaystyle{
\frac{a_1}{a_2}=
\frac{3 \sinh(2 \kkone t)-6 \kkone (t-t_0)}
 {6 \sinh \kkone t -2 \sinh^3(\kkone t) - 6 \kkone (t-t_0) \cosh \kkone t)},\ 
\kkone^2=c d,
}\\ \displaystyle{
w=\frac{3 \kkone}{2 k^2}
\frac{[\cosh \kkone t -\cosh k x][a_1 \cosh k x + a_2]}
 {\sinh \kkone t [a_1 \cosh \kkone t + a_2] }.
}
\end{array}
\right.
\label{eqsoltruncwnc43b}
\end{eqnarray} 
Each of these two new solutions is outside the class (\ref{eq-GP-class})
and depends on a single arbitrary constant 
($k_0$ in the first one, $t_0$ in the second one).

\subsubsection{Case $b=-4/5$}

With the value (\ref{eqF145}) of $F_1$,
the function $F_0$, algebraic in the derivatives of $F_1$,
is necessarily an affine function of $\zeta$, $\wp'$ and $x-g$ \cite[\S 18.6]{AbramowitzStegun},
\begin{eqnarray}
& & F_0= R_0 + R_1 \wp' + R_2 \zeta + (x-g(t)) R_3,
\end{eqnarray}
whose coefficients $R_{j}$ are rational in $\wp(x-g(t))$.
One then proves that, since $d$ is nonzero,
the discri\-minant $g_2^3-27 g_3^2$ must vanish,
thus reducing $F_1$ and $F_0$ to simply periodic functions,
\begin{eqnarray}
& & 
F_1=-\partial_x \psi(x,t),\ \psi=(k \tanh(k(x-g(t))))^2-(2/3) k^2 - \ex,\ k^2= 3 \varepsilon\ex,\ \varepsilon^2=1, 
\nonumber\\ & & 
F_0= -\frac{5}{6} g' \left[1+ (l_3-1/8) \ex^3 \psi^3\right],\ (\varepsilon-1) l_3=0.
\end{eqnarray}
Equation $E_1=0$ then generates the constraints
\begin{eqnarray}
& & {\hskip -10.0truemm}
\varepsilon=1,\ {g'}^2=\frac{36}{5} d,
\end{eqnarray}
thus restricting this solution to the particular case $b=-4/5$ of (\ref{eqF1tanh2}).

\subsubsection{Case $b=-1/2$}

Since we could not solve the system (\ref{eq-one-family-truncation-f-nonchar}),
we only solve here its particular case (\ref{eqTrunc1system}).
One first establishes the particular solution $F_0$ of $E_2=0$,
namely,
in the elliptic subcase $g_2^3-27 g_3^2\not=0$,
\begin{eqnarray}
& & 
y=\wp(x-g(t))-\ex,\
F_0=\frac{2 y^3 + g_3 + 8 \ex^3}{y^3}
\left[ \frac{g'}{3} - \frac{3 g_3'}{2 \Delta} \left((x-g) g_3-8 \ex^2  \zeta \right)\right]
\nonumber\\ & & \phantom{1234}
-\frac{3 g_3'}{2 \Delta} (8 \ex^2 y^2 -(g_3+8 \ex^3) (y -2 \ex)) \wp',\
\Delta= 27 (64 \ex^6-g_3^2),\
\end{eqnarray} 
and in the trigonometric subcase $g_2^3-27 g_3^2=0$, 
\begin{eqnarray}
& & 
F_1=\frac{1}{2}\partial_x \psi(x,t),\ \psi=(k \tanh(k(x-g(t))))^2-(2/3) k^2 - \ex,\ k^2=\pm 3 \ex, 
\nonumber\\ & & 
F_0= \frac{2}{3} g' \left[1+ \frac{4(1-\varepsilon)\ex^3}{\psi^3}\right].
\end{eqnarray}

Equation $E_1=0$ then generates the constraints
$g'=0$ in the elliptic subcase,
and $\varepsilon=-1$, ${g'}^2=9 d/2$ in the trigonometric subcase,
restricting the solution to (\ref{eqF1tanh2}) with $b=-1/2$.

\subsubsection{Case $b=-2/3$}

This case also happens to be a particular case of (\ref{eqF1tanh2}).


\section{Conclusion}

By a systematic investigation,
we have obtained several new solutions of this diffusion problem.
Two of them (section \ref{section-second-tanh}, Eq.~(\ref{eq-w-sol-odeg0}))
match all the physical constraints $b>0,c>0,d>0$
and
can serve to calibrate and validate the numerical schemes.

The two other new solutions,
Eqs.~(\ref{eqsoltruncwnc43a}) and
     (\ref{eqsoltruncwnc43b}),
only valid for a negative value of $b$,
could be useful for other diffusion problems governed by (\ref{EDPw})
with $b<0$.

Finally, in this short paper we have only investigated
those solutions $w$ which take into account one of the two movable poles.
Taking account of both poles \textit{via}
the two-singular manifold method (see \cite[\S 3.2.4.2]{CMBook} and references therein)
should certainly generate additional solutions.

\textit{Remark}.
As suggested by the two new solutions
Eqs.~(\ref{eqsoltruncwnc43a}) and
     (\ref{eqsoltruncwnc43b}),
the class $w$ equal to a second degree polynomial in $\cosh(k x)$
with time-dependent coefficients could also generate
physically interesting solutions.

\section*{Acknowledgments}
The author gratefully acknowledges the support of LRC M\'eso
and is happy to thank B.-J.~Gr\'ea, A.~Llor and R.~Motte for suggesting this
interesting and challenging problem.

\vfill \eject 


\begin{thebibliography}{00}

\bibitem{AbramowitzStegun} M.~Abramowitz, I.~Stegun,
\textit{Handbook of mathematical functions},
Tenth printing (Dover, New York, 1972).

\bibitem{Aronson} D.G.~Aronson, 
Regularity properties of flows through porous media,
SIAM J.~Appl.~Math.~{\bf 17} (2) (1969), 461--467.
http://www.jstor.org/stable/2099578

\bibitem{BureauMI} F.J.~Bureau,
Differential equations with fixed critical points,
Annali di Mat.~pura ed applicata {\bf LXIV} (1964) 229--364.

\bibitem{Conte1989} R.~Conte,
Invariant Painlev\'e analysis of partial differential equations,
Phys.~Lett.~A {\bf 140} (1989) 383--390.
https://doi.org/10.1016/0375-9601(89)90072-8.

\bibitem{CetraroConte} R.~Conte,
Exact solutions of nonlinear partial differential equations
by singularity analysis,
\textit{Direct and inverse methods in nonlinear evolution equations},
1--83,
ed.~A.~Greco,
Lecture notes in physics {\bf 632} (Springer Verlag, Berlin, 2003).
http://arXiv.org/abs/nlin.SI/0009024 
CIME school, Cetraro, 5--12 September 1999.

\bibitem{CMBook} R.~Conte and M.~Musette,
{\it The Painlev\'e handbook} (Springer, Berlin, 2008).
Second edition to appear (Springer, Switzerland, 2020).
\hfill\break\noindent
http://www.springer.com/physics/book/978-1-4020-8490-4 

\bibitem{GP} V.A.~Galaktionov and S.A.~Posashkov, 
Exact solutions and invariant spaces for nonlinear gradient diffusion equations,
Computational mathematics and mathematical physics {\bf 34} (1994) 313--321. 

\bibitem{GambierThese} B.~Gambier,                                  
Sur les \'equations diff\'erentielles du second ordre et du premier degr\'e
dont l'int\'egrale g\'en\'erale est \`a points critiques fixes,
Acta Math.~{\bf 33} (1910) 1--55.

\bibitem{LemmerLeach} R.L.~Lemmer and P.G.L.~Leach,   
The Painlev\'e test, hidden symmetries and the equation
$ y'' + y y' + k y^3 = 0$,
J.~Phys.~A {\bf 26} (1993) 5017--5024.

\bibitem{PHM2001} Pierre-Henri Maire,
\'Etude d'une \'equation de diffusion non-lin\'eaire.
Application \`a la discr\'etisation de l'\'equation d'\'energie cin\'etique
turbulente pour un mod\`ele de turbulence \`a une \'equation,
81 pages,
Rapport CEA D01 03661 (2001).

\bibitem{PaiBSMF} P.~Painlev\'e,
M\'emoire sur les \'equations diff\'erentielles dont l'int\'egrale
g\'e\-n\'e\-ra\-le est uniforme,
Bull.~Soc.~Math.~France {\bf 28} (1900) 201--261.

\bibitem{Zeldovich_Raizer} Y.~B.~Zel'dovich and Y.P.~Raizer,              
Physics of shock waves and high-temperature hydrodynamic phenomena, 
Vol.~2 (Academic press, New York, 1967).

\end{thebibliography}
\end{document}